# A Tool for Automatically Suggesting Source-Code Optimizations for Complex GPU Kernels

Saeed Taheri
Department of Computer Science
University of Utah
staheri@cs.utah.edu

Apan Qasem
Department of Computer Science
Texas State University
apan@txstate.edu

Martin Burtscher
Department of Computer Science
Texas State University
burtscher@txstate.edu

## Abstract

Future computing systems, from handhelds to supercomputers, will undoubtedly be more parallel and heterogeneous than today's systems to provide more performance and energy efficiency. Thus, GPUs are increasingly being used to accelerate general-purpose applications, including applications with data-dependent, irregular control flow and memory access patterns. However, the growing complexity, exposed memory hierarchy, incoherence, heterogeneity, and parallelism will make accelerator-based systems progressively more difficult to program. In the foreseeable future, the vast majority of programmers will no longer be able to extract additional performance or energy-savings from next-generation systems because the programming will be too difficult. Automatic performance analysis and optimization recommendation tools have the potential to avert this situation. They embody expert knowledge and make it available to software developers when needed. In this paper, we describe and evaluate such a tool. It quantifies performance characteristics of GPU code through profiling, employs machine learning models to estimate the suitability and benefit of several known source-code optimizations, ranks the optimizations, and suggests the most promising ones to the user if the expected speedup is sufficiently high.

## 1. Introduction

There are two primary difficulties with using accelerators such as GPUs. First, they can only execute certain types of programs efficiently, in particular programs with enough parallelism, data reuse, and regularity in their control flow and memory access patterns. Second, it is harder to write effective software for accelerators than for CPUs because of architectural disparities such as very wide parallelism, exposed memory hierarchies, lockstep execution, and memory access coalescing. Several new programming languages and extensions have been proposed to hide these aspects to various degrees and thus make it easier to program accelerators [1].

We study the alternative approach of making the programming and performance optimization easier for software developers who are not experts in GPU programming, specifically when it comes to complex irregular codes that are hard to parallelize. In particular, we describe a machine-learning-based recommendation tool for GPU kernels that automatically determines performance bottlenecks and suggests appropriate source-code optimizations, if any.

Several efficient GPU implementations of irregular algorithms have been published, showing that GPUs are capable of accelerating rather complex codes if they are implemented in a GPU-friendly fashion [2, 3, 4]. However, most software developers have no formal education in parallel programming, much less in accelerator programming, and could therefore greatly benefit from access to a performance/parallelism expert. Unfortunately, there are only relatively few such experts and each expert may only know a certain aspect or application domain. That raises the question of how to best deliver such expertise to programmers.

We believe the best solution to be automatic program analysis and recommendation tools. They embody the know-how of performance optimization experts and automatically determine where the bottlenecks lie and how to improve a given piece of code on a given system. Based on its analysis results, the tool recommends possible courses of action. Section 2 describes our tool in more detail.

Since the tool's recommendation accuracy hinges on how well it predicts the expected speedup of the optimizations in its database if they were applied to user-provided code, we evaluate it by comparing its predicted speedups with the actual speedups obtained when truly incorporating the suggested source-code optimizations. To make these comparisons possible, we wrote 64 versions each of two CUDA programs that include all possible combinations of six source-code optimizations and use different subset of these implementations to train and test our tool.

This paper makes the following contributions. 1) It describes how to build source-code recommendation tools that can automatically adapt to the underlying hardware and to changes in their optimization database. 2) We built such a tool for GPU programs and show that it delivers good recommendation accuracies on the platform and optimizations we tested, including on complex irregular CUDA code. 3) We study different scenarios to determine conditions that affect the tool's prediction accuracy.

The rest of this paper is organized as follows. Section 2 describes the design of our tool. Section 3 provides background information upon which the later sections are based. Section 4 discusses related tools and how they differ from our approach. Section 5 explains the experimental methodology. Section 6 presents the results and analyzes them. Section 7 concludes with a summary and future work.

## 2. Tool Design

Our tool employs a three-tiered design backed by an optimization database. The first tier performs code evaluation, the second tier analyzes the results, and the third tier handles the optimization selection. The tiers communicate through a simple interface. This makes it possible to design each tier independently and to replace any tier with an alternate implementation.

Tier 1 is concerned with evaluating code behavior and producing performance measurement data. We refer to these data as *feature vectors*. They are produced using NVIDIA's Visual Profiler [5]. It can measure a large number of hardware performance counter events such as instruction counts, cache hits/misses at different levels, etc. We normalize these *features* by the cycle count to make them independent of the runtime. The normalized features are then combined into a feature vector.

It should be noted that our tool does not depend on any particular profile information. Rather, the accuracy of the final recommendations simply improves with better profiling data. This makes the tool easy to port to platforms with different profilers or GPUs that support other performance counters.

Before we discuss the second tier, it is important to explain the content of the optimization database. The database is an unordered set of independent entries, where each entry represents an optimization, including a description with an example that illustrates how to apply it as well as pairs of before and after code samples that do not and do include the optimization, respectively. Each code sample includes one or more inputs to run it with.

A key feature of this database is that each entry is independent, making it easy to delete unwanted entries, modify existing entries, and add new entries. Thus, anybody can contribute optimizations, in particular experts from different domains. This makes the database very flexible, simple to port, and customizable to include only optimizations for a specific domain or hardware component.

Tier 2 analyzes the feature vector obtained from profiling the user's application to determine the most appropriate optimizations. Before it can do so, it must train itself on the before and after code samples from the database. It does this upon installation or when the database is modified by running the code samples through the Tier 1 profilers to obtain before and after feature vectors. From these vectors, it learns to recognize when a given optimization is needed and how much benefit it can deliver on the target platform. Tier 2 employs the machine learning algorithms listed in Section 3.4 for this purpose.

Tier 3 collects the recommendations from the second tier and sorts them by expected benefit. It then outputs the top choices if their benefit is above a preset threshold. The user can select how many recommendations to maximally display, whether to include the explanations and/or examples in the output, etc. These user-interface aspects are relatively straightforward and not the focus of this paper.

## 3. Background

### 3.1 GPU Architecture

This subsection provides a brief overview of the architectural characteristics of the Kepler-based Tesla K20c compute GPU we use and explains some of the features that make GPUs difficult to program. GPU programs require hierarchical parallelization across threads as well as across thread blocks of up to 1024 threads. The K20c consists of 13 streaming multiprocessors (SMs) to which the thread blocks are mapped. Each SM contains 192 processing elements (PEs) for executing the threads. Whereas each PE can run an individual thread of instructions, sets of 32 PEs are tightly coupled and must either execute the same instruction (operating on different data) in the same cycle or wait. This is tantamount to a SIMD instruction that conditionally operates on 32-element vectors. The corresponding sets of 32 coupled threads are called warps. Warps in which not all threads can execute the same instruction are subdivided by the hardware into sets of threads such that all threads in a set execute the same instruction. The individual sets are serially executed, which is called branch divergence, until they re-converge. To maximize performance, branch divergence has to be minimized, but it is typically difficult to implement programs in a manner such that sets of 32 threads follow the same control flow.

The memory subsystem is also built for warp-based processing. If the threads in a warp simultaneously access words in main memory that lie in the same aligned 128-byte segment, the hardware merges the 32 reads or writes into one coalesced memory transaction, which is as fast as accessing a single word. Warps accessing multiple 128-byte segments result in correspondingly many individual memory transactions that are executed serially. Hence, uncoalesced accesses are slower, but it is in general hard to write programs in such a way that sets of 32 threads access words from the same 128-byte segment. Part of the main memory, called constant memory, is reserved and can only be written by the CPU. GPU accesses to constant memory benefit from a special cache.

The PEs within an SM share a pool of threads called thread block, synchronization hardware, and a software-controlled data cache called shared memory. A warp can simultaneously access 32 words in shared memory as long as all words reside in different banks or all accesses within a bank request the same word. Barrier synchronization between the threads in an SM can take as little as a couple of cycles per warp. The SMs operate largely independently. They can only communicate through global memory (main memory in DRAM). The SMs support special instructions such as voting, where all threads in a warp compute a combined predicate (i.e., a reduction and broadcast operation), and rsqrtf, which quickly computes an approximation of one over square root. However, programmers may not be aware of the availability of such features, which can drastically boost the performance of code.

### 3.2 N-body Problem and Barnes-Hut Algorithm

To obtain test cases for evaluating our tool, we created 128 different versions of two *n*-body simulation codes (64 each) [6]. The first code, called NB, is regular and has $O(n^2)$ complexity. The second code, called BH, is irregular and has $O(n \log n)$ complexity. Both programs simulate the time evolution of a star cluster under gravitational forces for a given number of time steps. However, the underlying algorithm (see below) and the code base of the two implementations are completely different. *n* denotes the number of stars (aka bodies). Both of these codes have been written in such a way as there is essentially no execution taking place on the CPU.

The direct NB algorithm performs precise force calculations based on the $O(n^2)$ pairs of bodies. Since identical computations have to be performed for all bodies, the implementation is very regular and maps well to GPUs. The force calculations are independent and can be performed in parallel. In each time step, the $O(n^2)$ force calculation is followed by an $O(n)$ integration where each body's position and velocity are updated based on the computed force. For the values of *n* we consider, the integration represents an insignificant fraction of the overall execution time.

The Barnes-Hut (BH) algorithm approximates the forces acting on each body [7]. It recursively partitions the volume around the *n* bodies into successively smaller cells and records the resulting spatial hierarchy in an octree (the 3D equivalent of a binary tree). Each cell summarizes information about the bodies it contains. For cells that are sufficiently far away from a given body, the BH algorithm only performs one force calculation with the cell instead of one force calculation with each body inside the cell, which lowers the time complexity to $O(n \log n)$. However, different parts of the octree have to be traversed to compute the force acting on different bodies, making the control flow and memory-access patterns quite irregular. The force calculation is by far the most time consuming operation in BH, which is why we only consider source-code optimizations that affect this kernel. We use the BH implementation from the LonestarGPU suite [8]. It encompasses the algorithmic steps shown in Figure 1, each of which is implemented using one or multiple CUDA kernels. Since this implementation is irregular, we believe it is a good candidate for testing our tool.

```
bodySet = ...
foreach timestep do {            // O(n log n) + ordered sequential
  bounding_box = new Bounding_Box();
  foreach Body b in bodySet {    // O(n) parallel reduction
    bounding_box.include(b);
  }
  octree = new Octree(bounding_box);
  foreach Body b in bodySet {    // O(n log n) top-down tree building
    octree.Insert(b);
  }
  cellList = octree.CellsByLevel();
  foreach Cell c in cellList {   // O(n) + ordered bottom-up traversal
    c.Summarize();
  }
  foreach Body b in bodySet {    // O(n log n) fully parallel
    b.ComputeForce(octree);
  }
  foreach Body b in bodySet {    // O(n) fully parallel
    b.Advance();
  }
}
```

**Figure 1:** Pseudo code of Barnes-Hut algorithm

In summary, the NB code is relatively straightforward, has a high arithmetic intensity, regular control flow, and accesses memory in a strided fashion. In contrast, the BH code is quite complex (it repeatedly builds an unbalanced octree and performs various traversals on it), has a low arithmetic intensity, performs mostly pointer-chasing memory accesses, and has data-dependent control flow. Due to its lower time complexity, it is faster on a K20c GPU than the NB code when simulating more than about 15,000 stars.

### 3.3 Source-Code Optimizations

We modified our two test programs to make it possible to individually include or exclude all possible combinations of six source-code optimizations through conditional compilation, i.e., to produce 64 different versions of each programs. In particular, there are 32 versions of each program that do not and 32 that do include a particular source-code optimization. This enables us to create different subsets of these versions for training (providing before and after code samples), testing, and evaluating our tool.

For NB, we study the following six optimizations:

- CONST copies immutable kernel parameters (i.e., almost all of the parameters) once into the GPU's constant memory rather than passing them every time a kernel is called, i.e., it lowers the calling overhead.
- FTZ is a compiler flag that allows the GPU's floating-point ALUs to flush denormal numbers to zero, which results in faster computations. While strictly speaking not a code optimization, the same effect can be achieved by using appropriate intrinsic functions in the source code.
- PEEL separates the innermost loop of the force calculation into two consecutive loops, one of which has a known iteration count and can therefore presumably be better optimized by the compiler. The second loop performs the remaining iterations.
- RSQRT calls the CUDA intrinsic "rsqrtf()" to quickly compute one over square root instead of using the slower but slightly more precise "1.0f / sqrtf()" expression.
- SHMEM employs blocking, i.e., it preloads chunks of data into the shared memory, operates exclusively on this data, and then moves on to the next chunk. This reduces the number of global memory accesses.
- UNROLL uses a pragma to request unrolling of the innermost loop(s). Unrolling often allows the compiler to schedule instructions better and to eliminate redundancies, thus improving performance.

For BH, we study the following six optimizations.
- FTZ is identical to the corresponding NB counterpart.
- RSQRT is also identical to its NB counterpart.
- SORT approximately sorts the bodies by spatial distance to minimize the tree prefix that needs to be traversed during the force calculation.
- VOLA strategically copies some volatile variables into non-volatile variables and uses those in code regions where it is known (due to lockstep execution of threads in a warp) that no other thread can have updated the value. This optimization reduces memory accesses.
- VOTE uses thread voting instead of a shared-memory-based code sequence to perform 32-element reductions.
- WARP switches from a thread- to a warp-based implementation that is more efficient because it does not suffer from branch divergence and uses less memory as it records certain information per warp instead of per thread.

### 3.4 Machine Learning Algorithms

We utilize various subsets of the feature vectors from our test programs to train the Machine Learning (ML) methods in our tool such that they can learn how much speedup an optimization might provide under different conditions. The goal is to be able to predict by how much each of the optimizations in the database will improve or hurt the performance of a given CUDA kernel. Based on these predictions, the tool selects which optimizations to suggest.

Machine learning approaches generally use data attributes as features to perform classification/prediction. Each data entry can be viewed as a point in $N$-dimensional space, where $N$ is the number of attributes per data item. This allows, for example, to place each training data point into an $N$-dimensional space so that any test data point can be classified based on "nearby" training data points.

We examined three different ML approaches: linear and logistic regression, instance-based learners, and decision trees. Regression is concerned with modeling the relationship between variables that is iteratively refined using a measure of error in the predictions made by the model. Regression methods are important in statistics and have been cooped into statistical machine learning.

The instance-based learning model is a decision problem with instances or examples of training data that are deemed important to or required by the model. Such methods typically build a database of examples and compare new data to the database using a similarity measure to find the best match and make a prediction. The focus is on the representation of the stored instances and the similarity measures used between instances. In our experiments we use IBK, which is an instance-based classifier that uses the $k$-nearest neighbor (KNN) method for classification. During training, all labelled instances are recorded. When invoked on a new test instance, the model attempts to find the $k$ recorded instances that are most similar to the given test instance. Similarity is measured by the Euclidean distance between the feature vectors of the test and training instances. The mode value of the label for the *k* nearest neighbors is used to predict the outcome. Although we experimented with several different values of *k*, the results presented in this paper all use *k* = 10, which proved to be most effective.

Decision tree methods construct a model of decisions made based on the values of the attributes in the data. Decisions fork at each level in the tree until a leaf node is reached, where a prediction decision is made based on the training cases that reached the same leaf node. Decision trees are trained on data for classification and regression problems [9]. We employ M5P, a special type of decision-tree where each leaf node is a linear regression model. This model utilizes the M5 technique proposed by Quinlan [10]. First, an induction algorithm is used to construct a standard decision tree. Then a multivariate regression model is constructed for each node in the tree. However, instead of using all features in the regression model, only the features that appear in the subtree that contains the node are used. Finally, the leaf nodes in the tree are replaced with the newly constructed regression models. Once this regression-based decision tree has been built, standard pruning and smoothing techniques are applied.

## 4. Related Work

Paradyn [11] is one of the first tools for automatic performance analysis. It uses dynamic instrumentation to efficiently obtain performance profiles of unmodified executables. KOJAK [12], Scalasca [13, 14], Vampir [15] and VampirTrace [16] are trace-based tools that support MPI, OpenMP, and hybrid codes. For instance, the highly scalable Scalasca tool employs TAU's rich instrumentation capabilities [17] and processes the trace data in parallel. It scores and summarizes the trace report and shows it on a GUI.

Periscope [18] evaluates the performance while an application is running and searches for previously specified performance problems or properties. It is MPI-based and focused on efficient communication between cores/processors. TAU [17] is a portable tool for performance instrumentation, measurement, analysis, and visualization of large-scale parallel applications. Using the library wrapping benefit of TAU, TAUCuda [19] can measure GPU performance. It requires no modification of the source or binary code. The recently released Score-P tool [20] represents a portable infrastructure for performance measurement tools. Each of the above tools utilizes a different measurement output format. For example, the output format Vampir is OTF and the output format of Scalsca is EPILOG/CUBE. Score-P tries to integrate all of these tools into a unified measurement infrastructure. HPCToolkit [21, 22] generates statistical profiles using interval timers and hardware-counter interrupts and evaluates both application binaries and source code.

NVIDIA created tools such as the CUDA Performance Tools Interface (CUPTI) [23], Visual Profiler [24], and Nsight [25] that focus on GPU performance bottlenecks.

Some tools, such as PAPI CUDA [26] and VTune Amplifier XE [27], use hardware counters to measure the performance. eeClust [28] determines relationships between the behavior of parallel programs and the energy consumption of their execution.

Virtual Institute - High Productivity Supercomputing (VI-HPS) [29] is a collaboration of several partner institutions for improving the quality and accelerating the development process of complex simulation codes in science and engineering that are being designed to run on highly-parallel computer systems. Many well-known tools for parallel performance and measurement such as TAU, Scalasca and Vampir are designed and created by the partners of this big project. They also have a couple of ongoing and completed projects in the field of productivity and performance to improve their previous products. POINT, Score-P, SILC, HOPSA, PRIMA and LMAC are tools for integrating and improving the functionality of performance and measurement tools such as TAU and Vampir. For instance, LMAC adds the functionality of automatically examining performance dynamic for irregular behavior of parallel simulation codes to the established performance analysis tools Vampir, Scalasca, and Periscope.

Machine learning methods have also been used in MILEPOST GCC [30], a self-optimizing compiler that automatically learns the best optimization heuristics based on the behavior of the platform. There are also model-driven auto-tuning tools that are based on regression trees [31].

PerfExpert [32] is a tool that combines a simple user interface with an analysis engine to detect probable core-, socket-, and node-level performance bottlenecks in each important procedure and loop of a CPU application. For each bottleneck, PerfExpert provides a concise performance assessment. Unlike most of the tools described above, PerfExpert suggests steps that can be taken by the programmer to improve performance. In particular, its AutoSCOPE backend provides automatic recommendations for performance tuning, including compiler switches and optimization strategies with source-code examples [33]. It determines which suggestions to make by searching a manually annotated database of optimizations for the closest matches to PerfExpert's output metrics, which are derived from performance-counter measurements.

Our tool is most similar to that of PerfExpert/AutoSCOPE. We also use profiling based on hardware performance counters and compute derived metrics that are then used to identify suitable optimizations to recommend. However, instead of CPU procedures, we target complex GPU kernels, which can be challenging to make efficient. More importantly, instead of hand-annotating optimizations, which is tedious, error prone, and not very portable, our approach automates this step using ML algorithms that are trained using sample codes for each optimization. This not only makes it easy to port our tool to other systems but also enables the tool to automatically adapt the recommendations it makes to the performance characteristics of each system. Moreover, it provides the ability to alter the recommendation database without having to worry about how this change interacts with the remaining suggestions.

## 5. Experimental Methodology

We compiled the CUDA test programs using *nvcc* v6.0.1 with the *-O3 -arch=sm_35* flags. Our GPU is a Kepler-based 0.7 GHz Tesla K20c with 5 GB of main memory and 2496 CUDA cores distributed over 13 SMXs. Each SMX has 64 kB of fast memory that is split between the L1 data cache and the shared memory. The SMXs share a 1.5 MB L2 cache. For the machine learning methods, our tool leverages the algorithms implemented in Weka [34].

For the profiling, i.e., generating the feature vectors, we used *nvprof* from the Visual Profiler v6.5. We profiled each of the 128 versions of BH and NB described in Section 3.3 three times on the inputs shown in Table 1. Depending on the experiment, we use different subsets of the resulting feature vectors to train and test our tool. Table 2 lists the subsets used in each of the six experiments we performed. In experiments 1 through 4, we trained and tested based on the BH code. In experiments 5 and 6, the tool is trained on BH/NB and tested on NB/BH, respectively.

**Table 1:** Input sizes used for BH and NB

| NB | | BH | |
|---|---|---|---|
| Bodies | Time steps | Bodies | Time steps |
| 50,000 | 2 | 125,000 | 2 |
| 100,000 | 2 | 250,000 | 2 |
| 100,000 | 5 | 250,000 | 5 |
| 200,000 | 5 | 500,000 | 5 |
| - | - | 500,000 | 10 |
| - | - | 1,000,000 | 10 |

**Table 2:** Experiments for evaluating the speedup predictions

| Experiment | Training dataset | Training entries | Testing dataset | Testing entries | Testing includes training dataset | Train and test dataset from same prog. input |
|---|---|---|---|---|---|---|
| 1 | BH | 64 | BH | 192 | Yes | Yes |
| 2 | BH | 64 | BH | 128 | No | Yes |
| 3 | BH | 128 | BH | 64 | No | Yes |
| 4 | BH | 192 | BH | 64 | No | No |
| 5 | BH | 192 | NB | 64 | No | No |
| 6 | NB | 192 | BH | 64 | No | No |

Since the tool sorts its recommendations by predicted speedup, our evaluation focuses on comparing the actual speedup of the tested optimizations with the predicted speedup. If the predicted speedup is reasonably close to the actual speedup, our tool is able to suggest the most useful optimization(s) to improve performance.

The strategy we chose for evaluating and comparing the results after training the tool is the following. For each specific optimization, we removed all feature vectors from runs that included this optimization, which always leaves 32 feature vectors from runs that do not include the optimization. Testing these feature vectors on the trained tool generates six predicted speedups, one for each of the studied optimizations. The predicted speedup values are then compared to the actual (measured) speedup when truly including this optimization in the code.

The ratio of the actual speedup (AC) over the expected speedup (EX) shows how close the prediction is to the true speedup. If the predictions are accurate, the tool can use them to rank the optimization, i.e., suggest the most promising optimizations (if any) to the user based on the expected speedup. To enhance readability, we show the AC/EX ratios in strip charts. A strip chart plots the data along a line with each data point represented by a star. Note that the predictions do not have to be 100% accurate for our tool to work well. As long as the speedups are approximately correct, the tool will recommend the correct source-code optimizations, if any.

## 6. Results

This section presents the results of the prediction accuracy evaluations. We investigated three different machine learning methods to predict speedups: Logistic Regression, IBK, and M5P. Since the results of the logistic regression are substantially inferior to those of the other two methods, we only present results for IBK and M5P.

### 6.1 Train and Test on Same Code

When training and testing on the same program and input, the predictions are expected to be accurate. Instead of showing detailed strip charts for these simple experiments, we only compare the actual with the predicted speedup to see if they both show an increase or both show a decrease in performance. After all, if the predicted and the actual speedup are greater than one, it is correct for the recommendation tool to predict a performance gain. Similarly, if both the predicted and the actual speedup are less than one, using that optimization would hurt performance and not recommending the optimization is the correct behavior.

In experiment 1, we trained the tool based on the 64 feature vectors from a single run and input and tested all 192 feature vectors from the three runs of the same input, including the training data. Using the IBK method, on average over 97% of the predictions match the actual behavior, as shown in Table 3. The accuracy of the predicted behavior is significantly worse for M5P (86.4%). This reduced accuracy is largely due to M5P's inability to predict the behavior of FTZ, where it only achieves 57% accuracy. Upon further investigation we found that, in many of the test instances, FTZ applied by itself had very little impact on performance. Since M5P uses regression in the leaf nodes, even a small misprediction in the speedup can result in an incorrect final outcome (improvement vs. degradation). IBK does not suffer from this problem because it stores all of the training instances and is therefore able to predict the speedup of the training data exactly. IBK only enjoys this advantage if the training data include the test data. Next, we show how the accuracy of IBK is affected when we relax this assumption.

**Table 3:** Accuracy of negative/positive speedup predictions for different experiments and two ML methods

| Experiment | Accuracy IBK (%) | Accuracy M5P (%) |
|---|---|---|
| 1 | 97.3 | 86.4 |
| 2 | 96.0 | 86.4 |
| 3 | 96.3 | 33.3 |
| 4 | 92.0 | 81.6 |
| 5 | 83.6 | 33.3 |
| 6 | 55.7 | 60.1 |

#### 6.1.1 Non-overlapping Training and Test Data

In experiment 2, we trained on the 64 feature vectors of a single run and tested on the 128 feature vectors from the other two runs. Although the training data are not included in the testing data, we still expect high accuracy because all feature vectors stem from the same program running the same input multiple times. The IBK results (96%) are almost identical to experiment 1 with just a slight decrease in accuracy due to excluding the training data from the testing dataset. The results for M5P are also very similar to those of experiment 1. M5P uses just a few features, so excluding the training data does not affect its prediction accuracy much.

#### 6.1.2 Impact of Sample Size

In experiment 3, we trained the tool on 128 feature vectors and tested on the remaining 64 (experiment 2 uses the opposite approach). The expectation is that using more training data will improve the results. The prediction accuracies are comparable to the results from the previous experiments. Interestingly, the tested ML methods tend to underestimate the speedup. Nevertheless, the range of the ratios is 0.95 to 1.05 in all cases. These results show that adding more instances to the training data does not have a substantial impact on IBK. We note, however, that there is a significant drop in the accuracy of M5P. This is again explained by M5P's inability to accurately predict the behavior of FTZ. Although not shown here, the accuracy of M5P is much better in practice when using our tool with a threshold, i.e., when not recommending optimizations whose predicted speedup is below the threshold.

Obtaining about 96% prediction accuracy in the first three experiments is expected because training and testing on almost identical data (different runs of the same program and input) makes it easy for the tool to be accurate. In the following experiments, the training program input is different from the testing input.

### 6.1.3 Sensitivity to Program Input

In experiment 4, we trained the tool with all 192 feature vectors from the three runs on one program input and tested on 64 feature vectors each from the other program inputs. Figure 2 shows the results of the VOTE optimization with the IBK method. The Y axis of the chart shows the ratio of the Actual Speedup (AC) over the Expected Speedup (EX). The closer the data points are to 1.0 the more accurate the predictions are. The X axis represents different training and testing dataset combinations. Actually labeling the X axis resulted in illegible text, so we do not show the labels, which are not critical to the understanding of the paper. Note, however, that the input sizes increase from left to right and that the charts show sets of multiple strips for different runs of the same input size.

Most of the ratios in Figure 2 are around 1.0, meaning that the predicted speedups are close to the actual speedups. Unlike in the previous three experiments, where most of the IBK ratios were above 1.0, in this experiment the ratios are distributed quite evenly above and below the line. This is also true for the other optimizations shown in Figure 3. The few outliers in Figure 2 stem from test cases using the smallest inputs, which apparently result in poor feature vectors that throw off IBK.

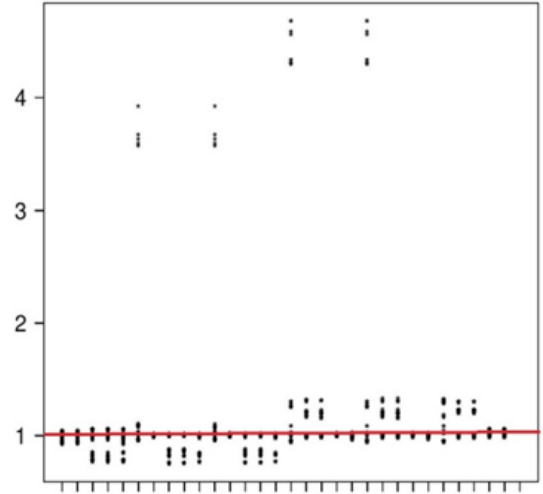

**Figure 2:** Ratios (AC/EX) of VOTE, Experiment 4, IBK

For the optimizations WARP, SORT, and VOLA shown in Figure 3, the predictions on smaller inputs are also less accurate. The plotted ratios are denser close to the 1.0 line for all three optimizations because of the higher accuracy with larger inputs.

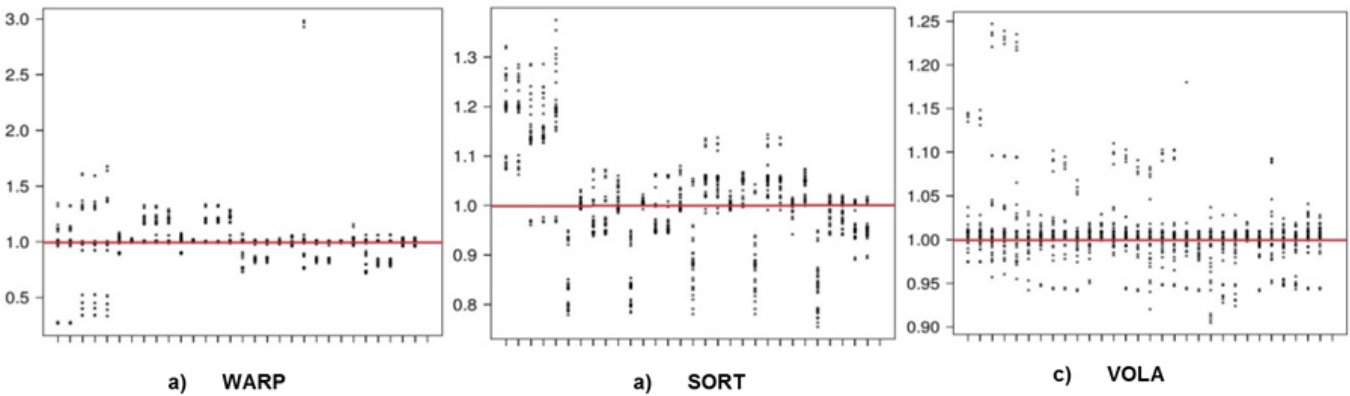

**Figure 3:** Ratios (AC/EX) of WARP, SORT, and VOLA, Experiment 4, IBK

Figures 4 and 5 show the IBK ratios for FTZ and RSQRT, respectively. The results are good for both FTZ and RSQRT. As shown in Table 3, the accuracy of positive/negative speedup is still 92% on average in experiment 4 for the IBK method. Clearly, training the tool on data from one input and testing on data from a different input does not hurt the tool's performance much. However, the accuracy of the prediction behavior of M5P is lower than IBK's (81.6%). This difference between absolute speedup prediction accuracy and behavior prediction accuracy, i.e., only predicting whether there will be a speedup, shows that the ratio of the actual speedup over the predicted speedup can be close to 1.0 yet the predicted speedup lies on the "other" side of the 1.0 line than the actual speedup. Fortunately, such cases are easily avoided in the recommendation tool by only suggesting optimizations that result in a speedup above the user-defined threshold.

### 6.2 Train and Test on Different Codes

Training the tool on a set of before and after feature vectors from code that is not related to the test code is the ultimate test of our approach (and the expected use case). In experiment 5, we trained on different versions of the BH code and used various versions of the NB code as test cases. In particular, this experiment shows results when we train the tool on data from an irregular GPU program and test it on a regular GPU program. Note that only the FTZ and RSQRT optimizations are common to both BH and NB. Hence, we can only compare the predicted and actual speedups of these two optimizations as we do not know the actual speedups of the remaining four BH optimizations when applied to NB.

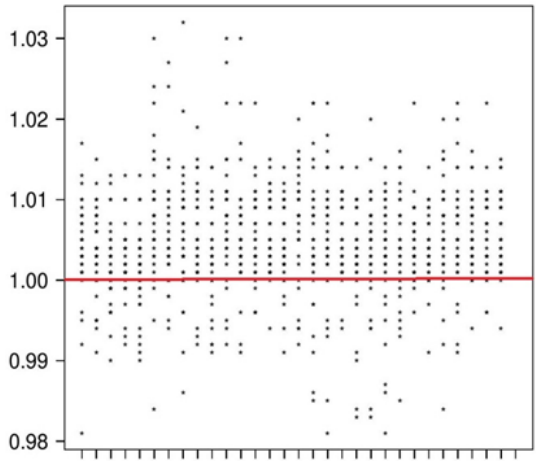

**Figure 4:** Ratios (AC/EX) of FTZ, Experiment 4, IBK

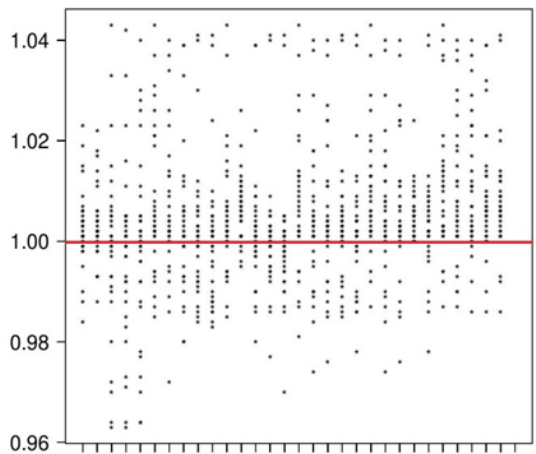

**Figure 5:** Ratios (AC/EX) of RSQRT, Experiment 4, IBK

Figures 6 and 7 show the results of experiment 5 using IBK. Almost half of the ratios are below the 1.0 line. The range of the ratios for FTZ is 0.2 to 1.7, which shows that the prediction accuracy of the speedup is not as close as it was in the previous experiments. For RSQRT, the ratios are spread even wider. As before, the prediction results for test cases with larger input sizes tend to be better. For each model, we tested all 64 feature vectors of each set of four inputs on the NB code.

Considering that we are training and testing on two different programs, the results are still good. The accuracy of the predictions for these two optimizations is almost 84%. The accuracy of the M5P method in this experiment for FTZ and RSQRT is only 33%. The reason for this low accuracy is that M5P uses very few features for making decisions. When the training and testing datasets stem from different programs, the possibility of accurate predictions based on just a few features is relatively low.

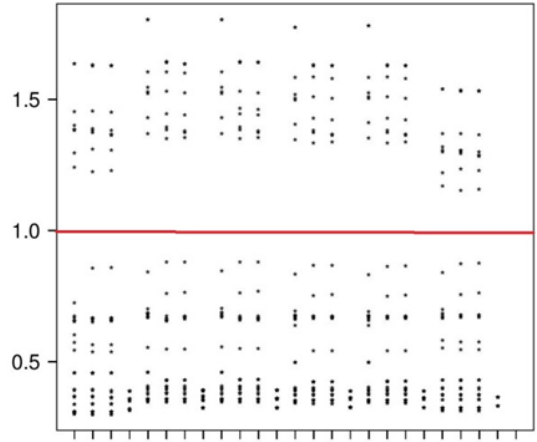

**Figure 6:** Ratios (AC/EX) of FTZ, Experiment 5, IBK

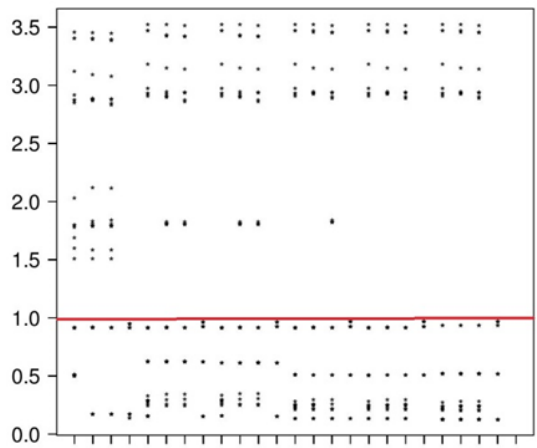

**Figure 7:** Ratios (AC/EX) of RSQRT, Experiment 5, IBK

Experiment 6 is identical to experiment 5 except we switched the training and testing datasets, that is, we trained the tool on the regular NB code and tested it on the irregular BH code. Interestingly, all of the predicted speedups for FTZ using the IBK method are lower than the actual speedups on the BH code as shown in Figure 8. RSQRT yields more accurate predictions as Figure 9 shows. The range of the ratios is 0.78 to 1.45 and most of the ratios are close to 1.0. For smaller training and testing inputs, the tool tends to overestimate the speedups.

Comparing the IBK results of experiment 6 with the corresponding results from experiment 5 in Table 3, we find that more accurate predictions are made when the tool is trained on irregular codes and tested on regular codes, which makes sense as irregular codes tend to be more complex.

In the first five experiments, the prediction accuracy of IBK is better than that of M5P. However, in experiment 6, the overall accuracy of M5P is better than that of IBK. Clearly, there is no ML model that is always the best for our tool. Apparently, M5P yields better performance because it narrows the features down to metrics that are significant for both irregular and regular codes. However, in most experiments, IBK yields more accurate speedup predictions than the other methods. Hence, IBK is the ML method of choice for our tool.

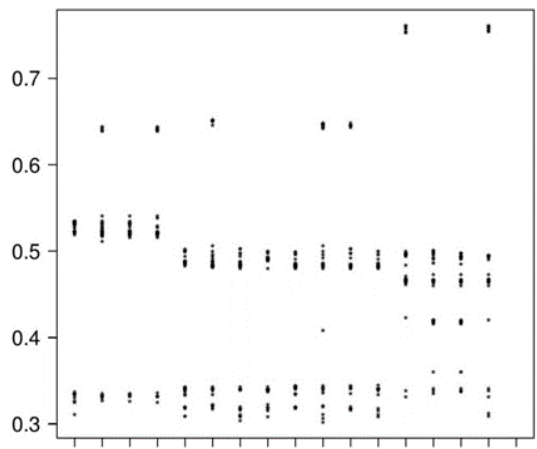

**Figure 8:** Ratios (AC/EX) of FTZ, Experiment 6, IBK

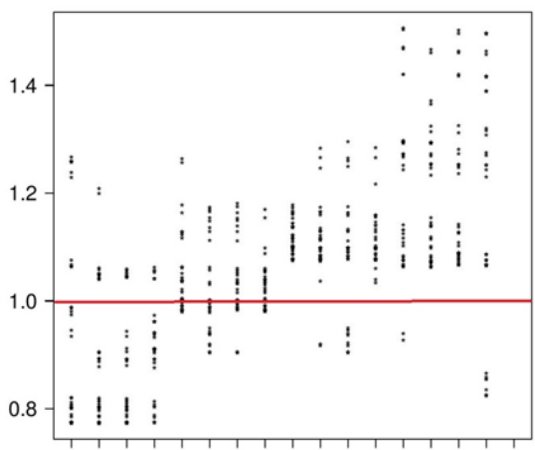

**Figure 9:** Ratios (AC/EX) of RSQRT, Experiment 6, IBK

## 7.  Summary and Conclusion

This paper describes and evaluates a tool to suggest source-code optimizations to programmers in order to improve the efficiency of their GPU code, including complex irregular codes. The tool needs to be trained on profile data from different code samples that do and do not include certain source-code optimizations. During this training, the tool builds machine-learning models for each optimization in its database so that it can later estimate the speedup for each optimization when presented with profile data from other programs. To measure and quantify the prediction accuracy, we profiled differently optimized GPU codes with multiple inputs to gather a large set of performance data. The tool ranks the optimizations based on the predicted speedup and suggests the top optimizations to the user if the predicted speedup is above a preset threshold. To evaluate the accuracy of the predicted speedups, we compared them to the actual speedups obtained when truly adding the respective source-code optimizations.

We performed six experiments of training models and predicting speedups. In the first four experiments, we trained and tested the tool on the BH code and obtained up to 97% prediction accuracy. In the remaining two experiments, where we train on BH/NB and test on NB/BH, the tool delivers up to 82% accuracy, i.e., most of the suggested source-code optimizations truly result in a speedup when they are implemented.

Based on the results from Section 5, the predictions of our tool are more precise when training on data obtained with larger program inputs. This makes sense as larger inputs result in more profiling data and more stable-state utilization of the GPU. Expectedly, training the tool with more data yields better predictions.

When training on code that is different from the tested code, we found that training based on irregular codes and testing on regular codes seems to result in better predictions than training on regular code and testing on irregular codes. This is likely a combination of two factors. First, regular codes are less complex, making them easier to predict in general. Second, the higher complexity of irregular codes probably provides more diverse training data, which yield better ML models for making the predictions.

We studied three different machine learning methods. Our results show that there is no clear winner. However, IBK generally performs very well when predicting the likely speedup of source-code optimizations. Hence, we use IBK in out tool.

We used differently optimized Barnes-Hut implementations as a representative irregular GPU code. Of course, using additional (irregular) codes for training would be better. Also, we studied six source-code optimizations. Larger numbers of optimizations can and should be used to better test the accuracy of our approach. To verify portability, our study should be repeated on additional types of GPUs. For the machine learning phase, we investigated three different methods. Other types of ML methods could, of course, also be employed for predicting the speedups.

## References

[1]  http://www.hpcwire.com/2014/01/09/future-accelerator-programming/

[2]  M. Mendez-Lojo, M. Burtscher, and K. Pingali. A GPU Implementation of Inclusion-based Points-to Analysis. 17th ACM SIGPLAN Symposium on Principles and Practice of Parallel Programming, pp. 107-116. February 2012.

[3]  Duane G. Merrill, Michael Garland, and Andrew S. Grimshaw. Scalable GPU Graph Traversal. 17th ACM SIGPLAN Symposium on Principles and Practice of Parallel Programming. February 2012.

[4]  M. Burtscher, R. Nasre, and K. Pingali. A Quantitative Study of Irregular Programs on GPUs. 2012 IEEE International Symposium on Workload Characterization, pp. 141-151. November 2012.

[5]  https://developer.nvidia.com/nvidia-visual-profiler

[6]  http://en.wikipedia.org/wiki/N-body_problem

[7]  http://en.wikipedia.org/wiki/Barnes%E2%80%93Hut_simulation

[8]  http://iss.ices.utexas.edu/?p=projects/galois/lonestargpu

[9]  http://machinelearningmastery.com/a-tour-of-machine-learning-algorithms/

[10] Ross J. Quinlan, Learning with Continuous Classes, Proceedings of the 5th Australian Joint Conference on Artificial Intelligence, Singapore, 343-348, 1992.

[11] B. P. Miller and J. K. Hollingsworth and M. D. Callaghan, The Paradyn Performance Tools and PVM, Proceedings of the Second Workshop on Environments and Tools for Parallel Scientific Computing: Townsend, TN, USA, 25–27 May

[11] 1994, pp. 201-210, Society for Industrial and Applied Mathematics, 1994.

[12] Bernd Mohr and Felix Wolf, KOJAK - a tool set for automatic performance analysis of parallel programs, Springer-Verlag, 2003.

[13] Zoltán Szebenyi, Brian J. N. Wylie, Felix Wolf: SCALASCA Parallel Performance Analyses of SPEC MPI2007 Applications. In Proc. of the 1st SPEC International Performance Evaluation Workshop (SIPEW), Darmstadt, Germany, volume 5119 of Lecture Notes in Computer Science, pages 99-123, Springer, June 2008.

[14] Markus Geimer, Felix Wolf, Brian J. N. Wylie, Erika Ábrahám, Daniel Becker, Bernd Mohr: The SCALASCA Performance Toolset Architecture. In Proc. of the International Workshop on Scalable Tools for High-End Computing (STHEC), Kos, Greece, pages 51–65, June 2008.

[15] W. E. Nagel and A. Arnold and M. Weber and H.-Ch. Hoppe and K. Solchenbach, VAMPIR: Visualization and Analysis of MPI Resources, 1996.

[16] Matthias S. Müller and Andreas Knüpfer and Matthias Jurenz and Matthias Lieber and Holger Brunst and Hartmut Mixand Wolfgang E. Nagel, Developing Scalable Applications with Vampir, VampirServer and VampirTrace, PARCO, Advances in Parallel Computing, Vol. 15, pp. 637-644, IOS Press, 2007.

[17] S. Shende and A. D. Malony, "The TAU Parallel Performance System," International Journal of High Performance Computing Applications, SAGE Publications, 20(2):287-331, Summer 2006

[18] Michael Gerndt and Karl Fürlinger and Edmond Kereku, Periscope: Advanced Techniques for Performance Analysis, PARCO, John von Neumann Institute for Computing Series, Vol. 33, pp. 15-26, Central Institute for Applied Mathematics, Jülich, Germany, 2005.

[19] Allen D. Malony, Scott Biersdorff, Wyatt Spear, and Shangkar Mayanglambam. 2010. An experimental approach to performance measurement of heterogeneous parallel applications using CUDA. In Proceedings of the 24th ACM International Conference on Supercomputing (ICS '10). ACM, New York, NY, USA, 127-136. DOI=10.1145/1810085.1810105 http://doi.acm.org/10.1145/1810085.1810105

[20] http://www.vi-hps.org/projects/score-p/

[21] Laksono Adhianto and Sinchan Banerjee and Michael Fagan and Mark Krentel and Gabriel Marin and John Mellor-Crummey and Nathan Tallent, HPCToolkit: Performance tools for parallel scientific computing, SC'08 USB Key, ACM/IEEE, November 2008.

[22] http://www.hpctoolkit.org

[23] https://developer.nvidia.com/cuda-profiling-tools-interface

[24] https://developer.nvidia.com/nvidia-visual-profiler

[25] http://www.nvidia.com/object/nsight.html

[26] Browne, S., Deane, C., Ho, G., Mucci, P. "PAPI: A Portable Interface to Hardware Performance Counters," Proceedings of Department of Defense HPCMP Users Group Conference, June, 1999.

[27] http://software.intel.com/en-us/intel-vtune-amplifier-xe

[28] Michael Knobloch and Timo Minartz and Daniel Molka and Stephan Krempel and Thomas Ludwig 0002 andBernd Mohr, Electronic poster: eeclust: energy-efficient cluster computing, SC Companion, pp. 99-100, ACM, 2011.

[29] http://www.vi-hps.org/projects/

[30] Fursin, Grigori, Cupertino Miranda, Olivier Temam, Mircea Namolaru, Elad Yom-Tov, Ayal Zaks, Bilha Mendelson et al. "MILEPOST GCC: machine learning based research compiler." In GCC Summit. 2008.

[31] Bergstra, J.; Pinto, N.; Cox, D., "Machine learning for predictive auto-tuning with boosted regression trees, "Innovative Parallel Computing (InPar), 2012, vol., no., pp. 13-14 May 2011.

[32] M. Burtscher, B.D. Kim, J. Diamond, J. McCalpin, L. Koesterke, and J. Browne. "PerfExpert: An Easy-to-Use Performance Diagnosis Tool for HPC Applications." SC 2010 Int. Conference for High-Performance Computing, Networking, Storage and Analysis. November 2010.

[33] Olalekan Sopeju, Martin Burtscher, Ashay Rane, and James Browne. AutoSCOPE: Automatic suggestions for code optimizations using PerfExpert. 2011 International Conference on Parallel and Distributed Processing Techniques and Applications, pages 19-25, July 2011.

[34] Mark Hall, Eibe Frank, Geoffrey Holmes, Bernhard Pfahringer, Peter Reutemann, Ian H. Witten, The WEKA Data Mining Software: An Update, SIGKDD Explorations, Volume 11, Issue 1, 2009.
9